\documentclass[12pt]{article}
\usepackage{a4,latexsym,graphicx,color,upgreek,amsmath}

\usepackage[numbers]{natbib}

\title{Characterisation and optimisation of foams for varicose vein sclerotherapy}
\author{T.G. Roberts$\,^1$ and S.J. Cox$\,^1$ and A.L. Lewis$\,^2$ and S.A. Jones$\,^2$\\ 
$\,^1$ Department of Mathematics, Aberystwyth University, SY23 3BZ, UK.\\
$\,^2$ Biocompatibles UK Ltd, a Boston Scientific Company,\\
Lakeview, Watchmoor Park, Camberley, Surrey, UK.}

\date{April 2020}

\begin{document}

\maketitle

\begin{abstract}
In this study we characterise the properties of foams used for varicose vein sclerotherapy. Their effectiveness is evaluated by predicting their yield stress and their flow profiles within a model of a vein. This information is represented using a Bingham number, which also takes into account the foam liquid fraction and the Sauter mean of the bubble size distribution. Based on this modelling, the most effective foams have a Bingham number $B \approx 600$ for a vein of diameter 2mm, in addition to a narrow bubble size distribution.
\end{abstract}

\section{Introduction}

Foam sclerotherapy is the process of using an aqueous foam to deliver surfactant (the sclerosant) to a varicose vein to damage vein wall endothelial cells, causing the vein to spasm, collapse and ultimately be re-absorbed into the body~\cite{coleridge09}. Foams with a broad range of properties are used in this treatment, with various methods of production, generally using the surfactants polidocanol or sodium tetradecyl sulphate. The physician administering the treatment may control the choice of gas, the bubble size (and its in-sample variation, which we refer to as polydispersity) and the liquid fraction of the foam, that is the proportion of liquid sclerosant present in a given volume of foam.

One of the reasons that a foam is used for this process is that the bubble microstructure endows it with beneficial flow properties. In particular, assuming complete vessel filling and no gravitational effects, these properties it to efficiently displace the blood in the vein, rather than to mix with it,  which would lead to deactivation of the sclerosant. In the language of (non-Newtonian) fluid dynamics, aqueous foams have a {\em yield stress}: when the foam is subjected to a large stress, it flows in the familiar manner of more common fluids such as water, but below a certain ``yield" stress $\uptau_0$ flow is arrested, and the foam is either stationary or moves as a plug. The stress acting on a foam within a vein varies widely, depending for example on the distance from the vein wall, and hence the size of the vein, and also the flow-rate.

Much of the literature on foam sclerotherapy concentrates on the properties of the foam before it enters the vein. 
The choice of gas affects foam stability: a foam created with carbon dioxide is much less stable than foam created with air~\cite{petersong10,beckitt2011air}, although it avoids the risks associated with introducing nitrogen into the cardiovascular system~\cite{carugoazzhohawl16}. The properties of the sclerosant influence the rate at which liquid drains from the foam (again reducing its lifetime)~\cite{carugoazzhohawl16,wollmann2010sclerosant,nastasa2015properties}, and choice of sclerosant is more significant than foam temperature and delivery rate~\cite{bai2018effect}. A foam with small bubbles and a narrow bubble size distribution offered high stability and cohesion in a biomimetic vein model, with consistent performance~\cite{carugo2015role}.


\begin{figure}
\centerline{
\includegraphics[width=0.6\textwidth]{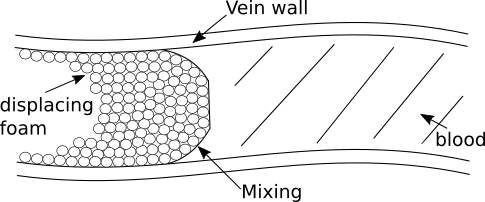}
}
\caption{The goal of the process of sclerotherapy is to entirely displace blood from a vein and then collapse the vein permanently. The shape of the front where the foam meets the blood is important in determining the degree of mixing and hence the efficacy of the process.}
\label{fig:setup}
\end{figure}

Let us consider a vein to be a straight cylinder with parallel walls (tortuosity can be introduced, but it does not change our argument) through which foam flows from some upstream injection point. The bubbles in the foam are packed closely together, and this induces an effective viscosity higher than that of the (continuous) liquid phase. Friction acts to slow down the foam close to the walls of the vein, and this induces a stress, which effectively liquifies the foam and allows it to flow (see figure~\ref{fig:setup}). Towards the centre of the vein, the effect of the walls is weaker, the stresses reduce, and a plug of rigid foam results.

This manifestation of the yield stress is what drives the process of sclerotherapy: the plug region in the centre of the vein displaces the blood in the vein, with little mixing, while to the sides the foam coats the vein wall with surfactant. Optimising the process of sclerotherapy requires a fairly high value of the yield stress. Too high, and the force required to push the foam out of a syringe and along a vein will be too great; too low, and the plug region will be too small, leading to excessive mixing of blood and foam close to the vein wall,  which hinders effective delivery of the sclerosant.

How should we characterize a foam so as to begin to optimize this process? The efficacy will depend upon the sclerosant chemistry, the properties of the foam itself, such as the bubble size, and on the properties of the vein into which the foam is to be delivered. For example, large veins might require a foam with slightly different properties to those required for small spider veins. In very small spider veins (telangiectasias, dilated interdermal venules $<1$ mm), liquid sclerosants are considered to be as effective as foam at displacing blood~\cite{zimmet2003sclerotherapy}.
The choice of gas, which we do not consider here, also affects surfactant transport within the foam~\cite{petersong10}.

Carugo {\it et al.}~\cite{carugoazzhohawl16} generated polidocanol foams in several different ways to compare the resulting bubble-size distribution and foam lifetime. The commercial product Varithena\textregistered \, ((polidocanol injectable foam 1\%, (also referred to as polidocanol endovenous microfoam, or PEM)), is generated from a device to produce a narrow bubble size distribution, and was compared in this evalulation to typical physician-compounded foams (PCFs) produced using both the Tessari method and the double syringe system (DSS). The DSS involves passing the liquid sclerosant and gas between two syringes joined by a simple straight connector. The Tessari method is nearly identical, but the straight connector is replaced by a three-way valve set at approximately 30$^{\circ}$. Despite the similarities between the methods, data suggests that the DSS method produces slightly better quality foam \cite{wollmann2010sclerosant}.

Typically, the liquid fraction of a PCF is high, between $0.2$ and $0.25$. Such foams have low yield stress and are likely to suffer from gravity override, in which the foam floats above the blood in a vein rather than displacing it. To effect their comparison, Carugo {\it et al.}~\cite{carugoazzhohawl16} produced foams with different methods but with a liquid fraction consistent with PEM, $\phi_l \approx 0.125$.

Our goal is to characterize the properties of foams used for sclerotherapy with the aim of improving the effectiveness and reproducibility of the process. In \S \ref{sec:yield} we describe our characterisation, in particular how the yield stress depends on bubble size and on any polydispersity in bubble size. In \S \ref{sec:Bing} we show how the yield stress affects the shape  of the displacement front within a vein, and draw conclusions about how this affects the efficacy of sclerotherapy in \S \ref{sec:concs}.

\section{Characterisation of bubble size distributions}
\label{sec:yield}

As the liquid fraction $\phi_l$ of a foam increases, the bubbles move apart. At $\phi_c \approx 0.36$, they are no longer deformed by any contact with their neighbours and are therefore spherical. Above this value, the foam is effectively a bubbly liquid, or a dilute suspension of bubbles, with zero yield stress. Wet foams, with a liquid fraction close to the critical value $\phi_c$ ($\phi_l \approx 0.2-0.25$ is typical of physician-compounded foams, or PCFs~\cite{star2018novel}) and will have a small yield stress and are likely to be inefficient in sclerotherapy.

\begin{figure}
    \centering
    \includegraphics[width=0.75\textwidth]{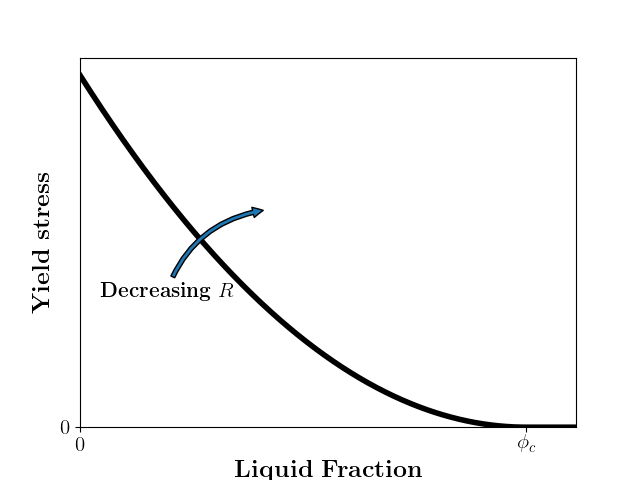}
    \caption{The yield stress $\uptau_0$ decreases with increasing liquid fraction $\phi_l$, cf. eqs.~(\ref{eq:phisquared}-\ref{eq:yieldstress}).
    As the bubble size $R$ decreases, the yield stress increases for $\phi_l$ less than the critical liquid fraction $\phi_c$, which remains fixed.
    }
    \label{fig:ys_vs_lf}
\end{figure}

The yield stress should therefore be described by some function $\uptau_0(\phi_l)$ that is positive for $\phi_l < \phi_c$ and reaches zero at this point. For vanishingly-small liquid fractions (``dry" foams), the yield stress is highest, recognising the consequences of a tight packing of polyhedral bubbles, making it difficult to deform the foam. The consensus points now to a dependence of the form~\cite{masonbw96,saintjalmesd99}:
\begin{equation}
\uptau_0 \sim \left(\phi_c - \phi_l \right)^2,
\label{eq:phisquared}
\end{equation}
shown in figure \ref{fig:ys_vs_lf}.
The squared dependence on the liquid fraction means that small differences in liquid fraction may have a disproportionately large effect. 

Bubble size also plays a role in determining the yield stress. At fixed liquid fraction, an appropriate scale for the stress in a foam is given by the Laplace pressure~\cite{mousse13}, the ratio of surface tension $\gamma$ to bubble size $R$. That is, for a given volume of foam, there are more interfaces when the bubbles are smaller, so the stress will increase. Supplementing eq.~(\ref{eq:phisquared}) with this scale suggests
\begin{equation}
\uptau_0 \approx 0.5 \frac{\gamma}{R} \left(\phi_c - \phi_l \right)^2,
\label{eq:yieldstress}
\end{equation}
where the pre-factor of $0.5$ brings this expression into close agreement with experimental data on foams and compressed emulsions \cite{masonbw96,princenk89}.

The mean bubble size $R$ is an average over the foam, i.e. the sum of the bubble radii divided by the number of bubbles. Instead of this unweighted mean, Princen and Kiss~\cite{princenk89} found that the Sauter mean radius, $R_{32}$,  is the most appropriate average bubble size for predicting the rheological properties of foams. The Sauter mean radius approximates the average bubble size based on the ratio of volume to surface area. Since in slow flows the response of a foam to deformation is dominated by the elasticity of the bubble interfaces, it should perhaps be no surprise the Sauter mean radius is a better measure of foam response. Indeed, Rouyer et al \cite{rouyerch05} state that their rheological data collapses on to a master curve {\em only} if the Sauter mean radius is used to scale the yield stress. The clear corollary is that bubble size polydispersity {\em does} affect the foam yield stress. We argue that it therefore influences the process of foam sclerotherapy.

In particular, the Sauter mean radius is sensitive to the presence of any large bubbles, recognising that a long tail in the bubble size distribution has a significant effect on foam response. It is for this reason that a narrow bubble size distribution is more appropriate for sclerotherapy.

In a foam of $N$ bubbles with different radii, the mean radius is $R = \langle R_b \rangle = \frac{1}{N} \sum R_b$. The Sauter mean radius, on the other hand, is proportional to the ratio of bubble volume to surface area, $R_{32} =  \langle R_b^3 \rangle / \langle R_b^2 \rangle$. The Sauter mean radius can be up to about 20\% greater than the mean radius in a disordered polydisperse foam~\cite{mousse13,feitosad08}.

\begin{figure}
\centerline{
\includegraphics[width=0.6\textwidth]{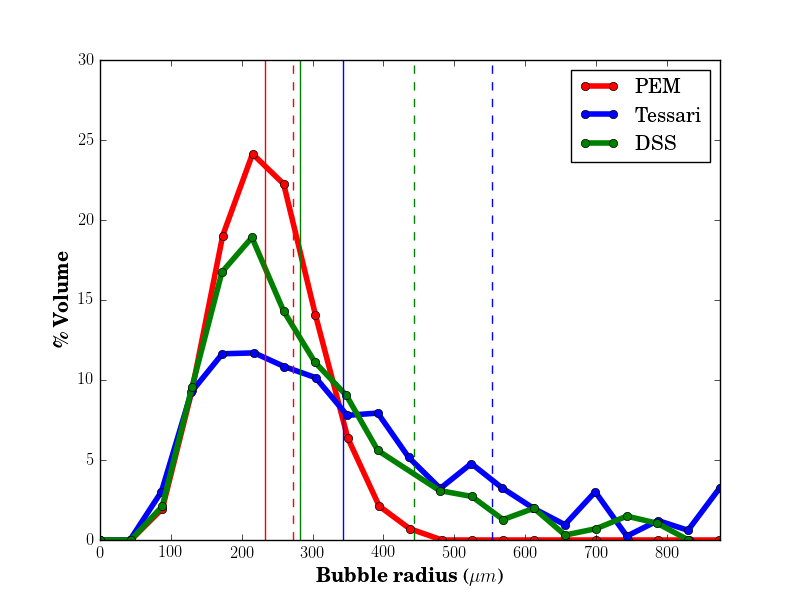}
}
\caption{The bubble distribution for the PEM, Tessari and DSS samples produced by Carugo {\it et al.}~\cite{carugoazzhohawl16}. The measurements are recorded $115$s after foam preparation in foams of the same liquid fraction. Solid vertical lines denote the mean bubble size ($R$) and dotted lines denote the Sauter mean radius ($R_{32}$).}
\label{fig:bubble_dist}	
\end{figure}

We calculate $R_{32}$ for the data from~\cite{carugoazzhohawl16} to determine the effect of how polydispersity is calculated on the expected value of the yield stress. Figure \ref{fig:bubble_dist} shows the bubble distributions of the three foams. For both the Tessari and DSS foams, there are several large bubbles with bubble radius greater than $500\mu m$; these are not present in the PEM foam. These large bubbles have a significant effect on the value of the Sauter mean radius of the foam, increasing it by $60\%$ and $56\%$ for the Tessari and DSS foams, respectively, over the usual mean, as shown in Table \ref{tab:yield_stress}.

This increase in the effective average bubble radius leads (cf. eq.~(\ref{eq:yieldstress}) with $R = R_{32}$) to a significant decrease in the predicted yield-stress of the foam. Assuming that the surface tension of all three foams are similar, about $\gamma = 30 \times 10^{-3}N/m$, allows us to calculate a value for the yield stress, as shown in Table \ref{tab:yield_stress}.

\begin{table}
\begin{center}
\begin{tabular}{c||c||c|c|c|c|c}
{Foam type} & $R$ ($\mu m$) & \textbf{$R_{32}$ ($\mu m$)}  & $\uptau_0$ (Pa) & $B$ \\
\hline
PEM & 233 & {272} & {3.04}  & {304.46} \\
Tessari & 343 & {553}  & {1.50} &  {149.92} \\
DSS & 283 & {443}  & {1.87}  & {187.19} \\
\end{tabular}
\caption{Measurements extracted from the data for PEM, Tessari and DSS foams from \cite{carugoazzhohawl16}: the mean and Sauter mean bubble size, the predicted value of the yield stress $\uptau_0$ (eq.~(\ref{eq:yieldstress})) and the Bingham number $B$ (eq.~(\ref{eq:bingeq})) using the Sauter mean radius. Using $R$ instead of $R_{32}$ to calculate $\uptau_0$ gives greater values but in the narrower range $2.4-3.6$Pa.}
\label{tab:yield_stress}
\end{center}
\end{table}

The table shows that the predicted yield stress of a foam depends strongly on the way in which the mean bubble size is calculated. Replacing the standard mean with the Sauter mean can reduce the predicted yield stress by up to one third. Since a high yield stress is important for sclerotherapy, this finding suggests that the efficacy of polydisperse foams may have been overestimated in the past.

We now turn to the consequences of this difference in yield stress for the degree to which the foam can effectively displace blood from a varicose vein.

\section{Characterisation of foams in veins: the piston effect}
\label{sec:Bing}

Having analysed foam properties, we now consider the effects of vein size and injection speed. The Bingham number, $B$, is a dimensionless measure (that is, it is universal, in the sense that it has no units) of the importance of a fluid's yield stress relative to the viscous stresses induced in the fluid by the flow. As we describe below, it is advantageous to use $B$ in place of $\uptau_0$ to represent the ``piston" effect  of a foam in a vein, i.e. how good it is at displacing fluid (blood) rather than mixing with it.

We consider a straight cylindrical vein of diameter $D$ and assume that fluid flows through it due to a difference in pressure between the injection site and the next (working) valve some distance along the vein. The pressure gradient  $G$ in the vein is the difference in pressure divided by this distance. It is dictated largely by, on the one hand, the force that can be used to depress the syringe without destroying the foam and, on the other, the need to deliver the foam before it starts to disintegrate. This disintegration occurs through diffusion-driven coarsening, in which the bubbles lose their gas to the surroundings~\cite{mousse13}.
The rate at which this coarsening occurs is determined by the solubility of the gas used to make the foam~\cite{petersong10}: faster for carbon dioxide, slower for nitrogen with oxygen presumable somewhere in between depending on the physiological environment.

In clinical delivery, 5ml of foam is injected in about 75 seconds. This corresponds to a flow rate $Q$ of roughly $6\times 10^{-5} {\rm m}^3/{\rm s}$. In a cylindrical vein the flow-rate is proportional to the applied pressure gradient $G$ and vein size cubed, $Q \propto G D^3$. Thus we estimate the pressure gradient to be $G \approx  Q/D^3$ which for a vein of diameter 2mm is of order $10^4$ Pa/m.

\begin{figure}
\centerline{
\includegraphics[width=0.6\textwidth]{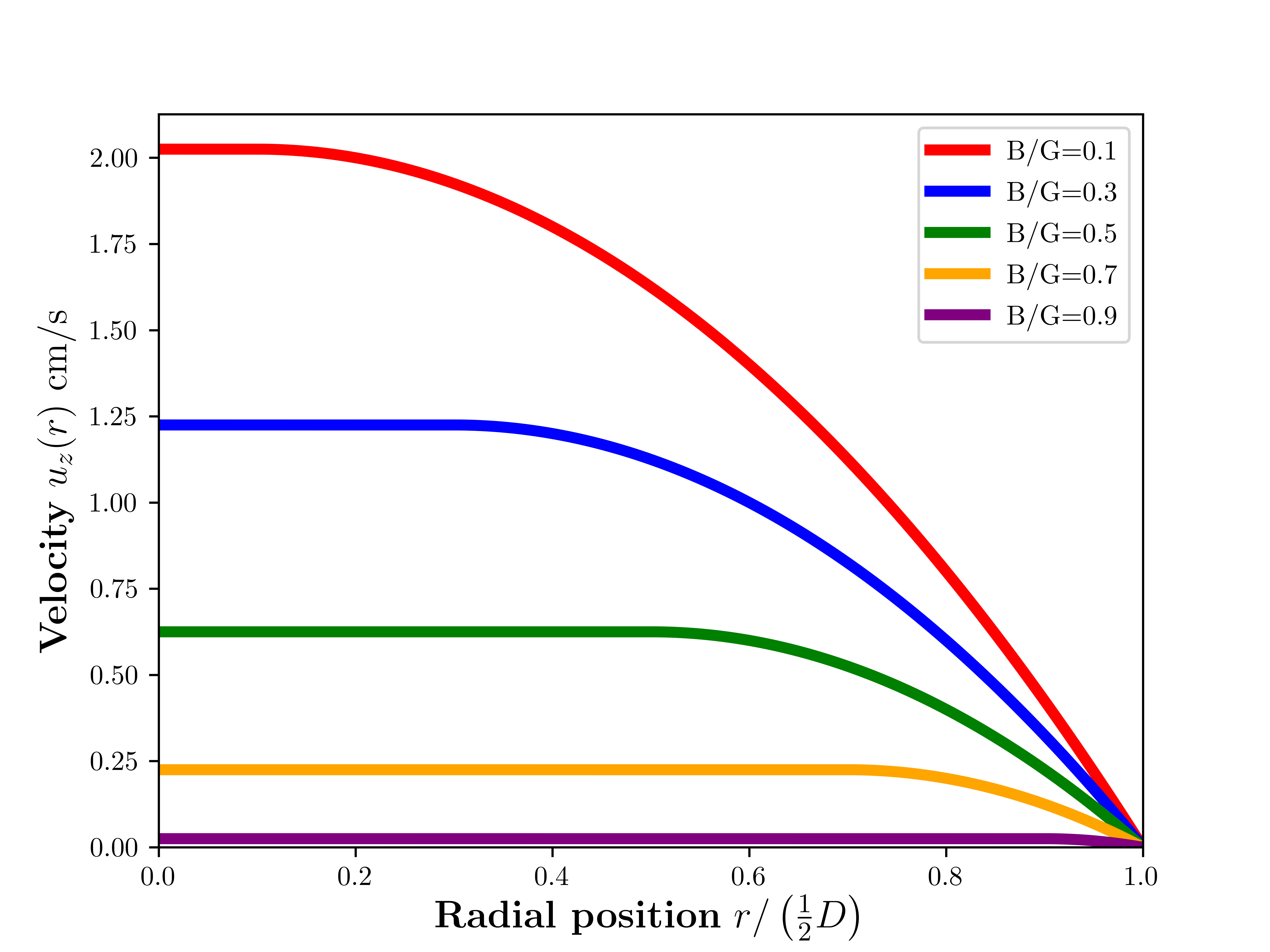}
}
\caption{Examples of the velocity profile for flow along a cylindrical vein for different values of the Bingham number $B$ relative to a fixed pressure gradient $G$.}
\label{fig:Bvelocity}
\end{figure}

We assume that there is sufficient friction at the vein wall to induce a no-slip boundary condition there. 
The Stokes equations for the slow flow of a yield stress (``Bingham") fluid in a cylindrical vein \cite{birddy83}  provide the distribution of the axial fluid velocity across the vein.
Examples of these velocity profiles are shown in figure~\ref{fig:Bvelocity}. 

Using $r$ to represent radial position in the vein, we note that there is an interface at $r=r_0$ which separates the plug region in the centre of the vein, with constant velocity, from the yielded region close to the walls. It is this plug region that is effective in displacing blood. Its size is directly proportional to the Bingham number, $r_0 = B/G$. A smaller pressure gradient $G$ (and hence a slower flow), or a fluid with high Bingham number $B$ (and hence yield stress), will have a wide plug of foam pushing down the vein. Thus optimisation of $B$ is necessary.

\begin{figure}
\centerline{
\includegraphics[width=0.6\textwidth]{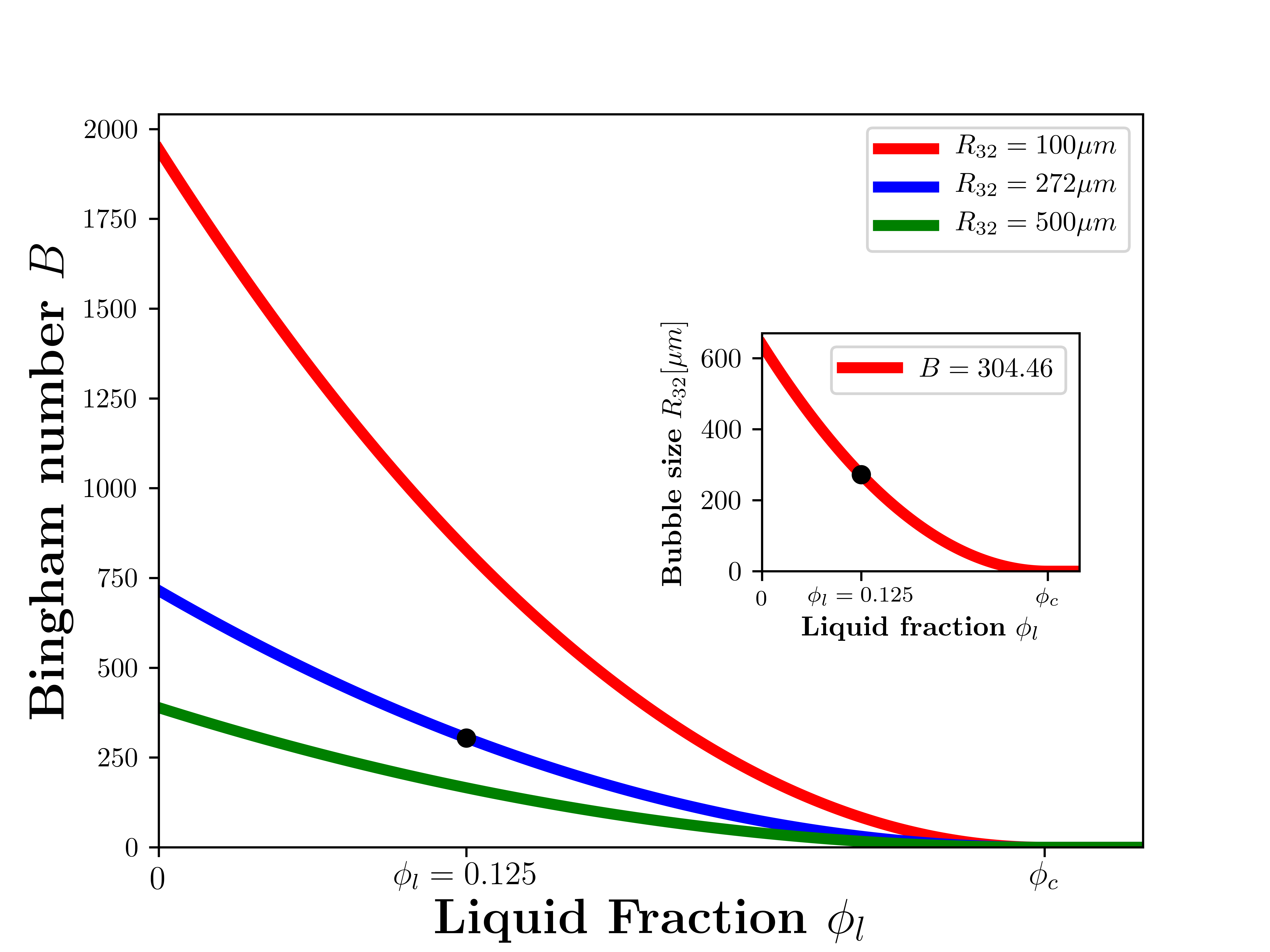}
}
\caption{Representative values of the Bingham number $B$, showing the strong dependence on the liquid fraction of the foam and the bubble size. Lines are shown for three values of the Sauter mean bubble radius for a vein with diameter 2mm, with fluid speed fixed at $U = 1$ cm/s, viscosity $\mu = 1\times 10^{-3}$ Pa s  and interfacial tension $\gamma = 30\times 10^{-3} $N/m. The values for PEM  are shown with a black dot. The inset shows the relationship between the bubble size $R_{32}$ and the liquid fraction $\phi_l$ that will result in the same value of the Bingham number $B$.}
\label{fig:Bphi}
\end{figure}

The Bingham number depends not only on the foam properties, but also on the flow itself. If the sclerosant solution forming the liquid phase of the foam has viscosity $\mu$ (of the order of $10^{-3}$  Pa s, usually significantly below the effective viscosity of the foam) and flows with speed $U$, of order $\frac{1}{2}G R^2$, 
then we write
\begin{equation}
B \;=\; \uptau_0 \frac{ D }{\mu U }  \;= \;  0.5\frac{\gamma D }{\mu R_{32} U} \left(\phi_c - \phi_l \right)^2,
\label{eq:bingeq}
\end{equation}
using eq.~(\ref{eq:yieldstress}). This relationship is shown in figure \ref{fig:Bphi} for different values of the bubble size. We advocate that $B$ should be used to characterize the process of foam schlerotherapy.

We can now use eq. (\ref{eq:bingeq}) and the yield-stress values in Table \ref{tab:yield_stress} to find the respective values of the Bingham number for PEM, Tessari and DSS foams. As the Bingham number is also dependent on the flow properties, we assume a fixed fluid speed $U = 1$cm/s and a surfactant viscosity $\mu = 1\times 10^{-3}$ Pa s. This allows us to determine the values for the Bingham number shown in Table \ref{tab:yield_stress}. The data shows that the Bingham number of the PEM foam is double the value of the Bingham number for the Tessari foam and $62\%$ greater than the DSS foam. 

\begin{figure}
\centerline{
	\includegraphics[width=0.6\textwidth]{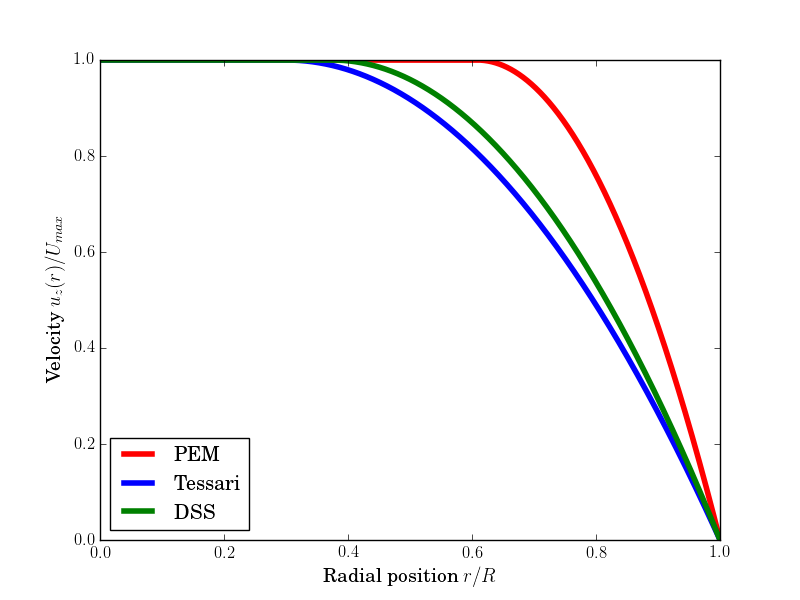}
}
\caption{For a constant pressure-gradient $G = 10^4$ Pa/m, we approximate the velocity profiles for PEM, Tessari and DSS foams flowing through veins of diameter of 2mm. We scale by the maximum velocity in each case.}
\label{fig:foamvelocity}	
\end{figure}

In practice, $G$ is set by the rate at which the surgeon injects the foam. The value of the pressure gradient needed to achieve a fluid speed of $1$cm/s is $G = 10^4$~Pa/m. 
Figure \ref{fig:foamvelocity} shows predicted velocity profiles in a cylindrical vein for each of the measured foams. Note how the width of the plug region is smaller for foams made with the Tessari method compared to the PEMs. Hence the difference in Bingham number is clearly correlated with a significant difference in the extent of the plug regions for the PEM and physician compounded foams.

\section{Conclusion}
\label{sec:concs}

We have described a way to characterize the properties of foams used for sclerotherapy and to evaluate their effectiveness by introducing a framework to predict their yield stress and flow profiles. We compared the effectiveness of a PEM and PCFs by considering the Sauter mean of their bubble size distributions~\cite{carugoazzhohawl16}. 

The Sauter mean $R_{32}$ is more greatly affected than the usual mean by the presence of large bubbles, which in turn affects a foam's yield stress. Given $R_{32}$, we estimate the value of the yield stress $\uptau_0$ for different foam liquid fractions $\phi_l$. Our approximation of $\uptau_0$ allows us to estimate the shape of the displacement front of foam within a straight vein for a given flow rate and vein diameter. We use the value of $\uptau_0$ to define the Bingham number $B$ (eq.~(\ref{eq:bingeq})) as a dimensionless measure of the ability of a foam to displace blood.

Our calculations suggest that an optimal foam should have yield stress close to $\uptau_0 = 3$ Pa (eq.~(\ref{eq:yieldstress})) and hence sclerotherapy treatments should aim for a Bingham number $B \approx 600$ for a vein of diameter $D \approx 2$mm, in addition to a narrow bubble size distribution. In larger veins, slightly larger bubbles will result in this same value of $B$ and {\it vice versa}.

The steepness of the curves in figure \ref{fig:Bphi} shows that such a value of $B$ may be difficult to obtain.
A foam with high liquid fraction, for example 0.25, typical of PCFs, would need tiny bubbles (of order tens of microns) to be effective at displacing blood. Such bubble sizes are not possible to obtain with e.g. the Tessari method~\cite{tessari2000}. On the other hand, a dry (low liquid fraction) foam, such as could be obtained by leaving a foam to drain before injection, would have bubbles that are approaching the width of the vein, and would therefore be ineffective.

Finally, we note that the affected vein should be kept as straight as possible during treatment: vein curvature induces additional stresses within the foam, leading to a greater degree of yielding and therefore reducing the size of the plug~\cite{roberts2020analytic}. In turn, this leads to a less effective displacement of blood and a greater chance of polidocanol deactivation due to mixing.

\section*{Acknowledgements}

We acknowledge financial support from the UK Engineering and Physical Sciences Research Council (EP/N002326/1) and a PhD studentship from BTG.

\bibliography{piston.bib}

\begin{thebibliography}{10}

\bibitem{coleridge09}
P.~Coleridge~Smith.
\newblock Rigidity percolation in particle-laden foams.
\newblock {\em Phlebologie}, 24:62--72, 2009.

\bibitem{petersong10}
J.D. Peterson and M.P. Goldman.
\newblock An investigation into the influence of various gases and
  concentrations of sclerosants on foam stability.
\newblock {\em Dermatologic surgery}, 37:12--17, 2001.

\bibitem{beckitt2011air}
T.~Beckitt, A.~Elstone, and S.~Ashley.
\newblock Air versus physiological gas for ultrasound guided foam sclerotherapy
  treatment of varicose veins.
\newblock {\em Eur. J. Vasc. Endovasc. Surg.}, 42:115--119, 2011.

\bibitem{carugoazzhohawl16}
D.~Carugo, D.N. Ankrett, X.~Zhao, X.~Zhang, M.~Hill, V.~O’Byrne, J.~Hoad,
  M.~Arif, D.D. Wright, and A.L. Lewis.
\newblock Benefits of polidocanol endovenous microfoam
  (varithena{\textregistered}) compared with physician-compounded foams.
\newblock {\em Phlebology}, 31:283--295, 2016.

\bibitem{wollmann2010sclerosant}
J.C. Wollmann.
\newblock Sclerosant foams: stabilities, physical properties and rheological
  behavior.
\newblock {\em Phlebologie}, 39:208--217, 2010.

\bibitem{nastasa2015properties}
V.~Nastasa, K.~Samaras, C.~Ampatzidis, T.D. Karapantsios, M.A. Trelles,
  J.~Moreno-Moraga, A.~Smarandache, and M.L. Pascu.
\newblock Properties of polidocanol foam in view of its use in sclerotherapy.
\newblock {\em Int. J. Pharm.}, 478:588--596, 2015.

\bibitem{bai2018effect}
T.~Bai, W.~Jiang, Y.~Chen, F.~Yan, Z.~Xu, and Y.~Fan.
\newblock Effect of multiple factors on foam stability in foam sclerotherapy.
\newblock {\em Sci. Rep.}, 8:1--7, 2018.

\bibitem{carugo2015role}
D.~Carugo, D.N. Ankrett, V.~O'Byrne, D.D. Wright, A.L. Lewis, M.~Hill, and
  X.~Zhang.
\newblock The role of clinically-relevant parameters on the cohesiveness of
  sclerosing foams in a biomimetic vein model.
\newblock {\em J. Mater. Sci.: Mater. Med.}, 26(11):258, 2015.

\bibitem{zimmet2003sclerotherapy}
S.E. Zimmet.
\newblock Sclerotherapy treatment of telangiectasias and varicose veins.
\newblock {\em Tech. Vasc. Interv. radiol.}, 6:116--120, 2003.

\bibitem{star2018novel}
P.~Star, D.E. Connor, and K.~Parsi.
\newblock Novel developments in foam sclerotherapy: Focus on
  varithena{\textregistered}(polidocanol endovenous microfoam) in the
  management of varicose veins.
\newblock {\em Phlebology}, 33(3):150--162, 2018.

\bibitem{masonbw96}
T.G. Mason, J.~Bibette, and D.A. Weitz.
\newblock Yielding and flow of monodisperse emulsions.
\newblock {\em J. Coll. Interf. Sci.}, 179:439--448, 1996.

\bibitem{saintjalmesd99}
A.~Saint-Jalmes and D.J. Durian.
\newblock Vanishing elasticity for wet foams: Equivalence with emulsions and
  role of polydispersity.
\newblock {\em J. Rheol.}, 43:1411--1422, 1999.

\bibitem{mousse13}
I.~Cantat, S.~Cohen-Addad, F.~Elias, F.and~Graner, R.~H{\"o}hler, O.~Pitois,
  F.~Rouyer, and A.~Saint-Jalmes.
\newblock {\em Foams: structure and dynamics}.
\newblock OUP Oxford, 2013.

\bibitem{princenk89}
H.M. Princen and A.D. Kiss.
\newblock Rheology of foams and highly concentrated emulsions: Iv. an
  experimental study of the shear viscosity and yield stress of concentrated
  emulsions.
\newblock {\em J. Coll Int. Sci.}, 128:176--187, 1989.

\bibitem{rouyerch05}
F.~Rouyer, S.~Cohen-Addad, and R.~H{\"o}hler.
\newblock Is the yield stress of aqueous foam a well-defined quantity?
\newblock {\em Coll. Surf. A}, 263:111--116, 2005.

\bibitem{feitosad08}
K.~Feitosa and D.J. Durian.
\newblock Gas and liquid transport in steady-state aqueous foam.
\newblock {\em Euro. Phys. J. E}, 26:309--316, 2008.

\bibitem{birddy83}
R.B. Bird, G.C. Dai, and B.J. Yarusso.
\newblock The rheology and flow of viscoplastic materials.
\newblock {\em Rev. Chem. Eng.}, 1:1--70, 1983.

\bibitem{tessari2000}
L.~Tessari.
\newblock New technique for obtaining sclero-foam.
\newblock {\em Phlebologie}, 53:129, 2000.

\bibitem{roberts2020analytic}
T.G. Roberts and S.J. Cox.
\newblock An analytic velocity profile for pressure-driven flow of a bingham
  fluid in a curved channel.
\newblock {\em J. non-Newt. Fl. Mech. (in press)}, 2020.

\end{thebibliography}
\bibliographystyle{unsrt}

\end{document}